\documentclass[aps,twocolumn,showpacs,superscriptaddress]{revtex4-1}
\usepackage{amsmath,amssymb}
\usepackage{graphics,graphicx}
\usepackage{dcolumn,bm}
\usepackage{psfrag}
\newcommand{\cu}
{\affiliation{Department of Physics, University of Calcutta, 
92 Acharya Prafulla Chandra Road, Kolkata 700009, India.}}

\begin{document}
 
\title{Damage spreading transition in an opinion dynamics model}

%\author{Soham Biswas}%
%\tifr
\author{Abdul Khaleque}%
\cu
\author{Parongama Sen}%
\cu

\begin{abstract}

 We study the damage spreading phenomena in two different ways in a opinion 
dynamics model introduced recently. This kinetic exchange type model is characterized
 by a fraction $q$ of negative interactions and shows the presence of an order-disorder
transition at $q_c$. In the traditional method, 
two replicas of the population  are considered in which the 
opinion of all the agents are identical initially except for a single agent. 
The systems are then allowed to evolve identically. 
In the other method, the initial opinions are identical for all agents but
the two replicas  are evolved independently. 
In both cases, a damage spreading transition occurs at $q_d$ where
$q_d \approx 0.18$ in the traditional method and $q_d =0$ for the other; 
the damage increases with $q$ above $q_d$ and attains a constant value for $q\geq q_c$.
However, the correlation between the evolved states 
above $q_c$ is clearly different in the two methods.

\end{abstract}

\pacs{05.45.-a,87.23.Ge,74.40.De}

\maketitle

\section{Introduction}
%%%%%%%%%%%%%%%%%%%%%%%%%%%%%%%%%%%%%%%%%%%%%%%%%%%%%%%%%%5
 In systems with randomness and disorder, damage spreading is an important dynamical study that was
 first introduced in the context of biologically motivated systems \cite{kauffman}.
Later, in physics, a number of studies were conducted in the Ising model
and cellular automata \cite{cruetz,stanley,coniglio,poole,stauffer,grassberger,wang,wang1,vojta,vojta1,vojta2,
glotzer,svenson,derrida-pom,derr-weis,derrida-stauffer,weisbuch,herrmann,silva,bagnoli,geza} in which two copies 
of the system were made different by a small amount  and were evolved using the same random numbers.  
The time development  of the ``damage" $D(t)$, which is a measure of the difference between the two replicas, 
is one of the important features  that is studied.  One expects the damage will reach a constant value at long times, i.e.  
$D(t \to \infty) = D_{sat}$ where $D_{sat}$ can be zero or finite. 
In case it is zero, the damage does not survive. Thus the  question 
whether  there could be a damage spreading transition, i.e.   whether  damage is nonzero only above a certain value of the 
driving parameter (which may be  temperature in  physical systems), becomes an important study also. 
Such a transition point may not necessarily coincide with the order-disorder transition 
point, if any.

Traditionally, in spin systems, damage spreading is  studied  in two ways. 
One can either let the system equilibrate and then make a slightly
damaged replica and study the evolution of both. Or, two slightly different  replicas 
may be allowed to evolve right from the beginning \cite{coniglio}. 
Important feature is, whichever way it is done, the replicas are
evolved identically in this so called traditional method (TM).

In some  recent works,  the fate of two identical copies
of the Ising model which were   allowed to evolve independently was studied
\cite{de-oliv,machta}. This is done to  study the effect of 
environment which can  introduce differences in 
two systems born with identical features. This has been termed as a ``nature versus nurture''
(NVN) phenomena \cite{machta}.   
In one and two dimensions, this leads to a power
law decay of the overlap between the two systems in time 
signifying that the copies become more and more `damaged' in time.
However, such studies have been limited to low temperatures. 
We use this as an alternative method of damage spreading study calling it the NVN method. 

We intend to study whether  the two methods lead to any qualitative and 
quantitative difference as far as damage spreading is concerned 
in a particular system. 
We choose an opinion dynamics model to study the damage spreading phenomenon
employing both the  methods. 
In contrast to spin models, where different dynamical algorithms (e.g., 
heat bath, metropolis etc.) may lead to qualitatively different results \cite{mariz,hinrichsen} 
as far as damage spreading is concerned,  
opinion dynamics models have well-defined dynamical rules.  
It is also important to study damage spreading in 
opinion dynamics models  to explore whether 
an initial small difference in opinion can induce drastic changes in the
opinions of all the agents  \cite{sociobook,castellano}.
Various  models of opinion dynamics  exist in the literature, 
although damage spreading has  been studied in comparatively less extent \cite{bern,fortu,guan}. 
The model studied in this work is relatively new \cite{biswas}, 
it contains a single parameter and shows  
an order-disorder transition. Thus here it is also possible to see whether the damage spreading transition, if any, occurs at  the
the known phase  transition point or not.
Secondly, 
the model  has the property that individual opinions may go on changing 
even when the global average opinion reaches a steady state, 
hence we expect that the time dependence of the damage itself 
may show interesting features.
The  model  has the additional advantage that it can be studied 
using both discrete and continuous opinions and the nature of  randomness 
here can also be modified. 
It therefore  gives us the opportunity to
study  different cases within a common framework.

In the next section, the model and the method are described. 
Results are presented in section III followed by a summary and discussion in the last section.

\section{Model and Method}
In this work, we study the opinion dynamics model proposed in \cite{biswas}, where opinions can be  modeled as discrete or continuous variables. 
Here two individuals modify their opinions by the so called ``kinetic exchange'' scheme \cite{sociobook,biswas2,anirban}.
The  opinions are subject to change due to the mutual binary interactions which can be both
positive as well as  negative. Let $O_{i}(t)$ be the opinion of the $i$th
agent at time $t$, then after an interaction of the $i$th and
$j$th agents, their opinions at time $t+1$ are changed
according to
\begin{equation}
\begin{split}
 O_{i}(t+1) = O_{i}(t) + {\mu_{ij}}O_{j}(t)\\
 O_{j}(t+1) = O_{j}(t) + {\mu_{ij}}O_{i}(t),
\end{split}
\end{equation}
where $\mu_{ij}$ is random, either $+1$ or $-1$. We consider both discrete ($O_{i}=0,\pm 1$) 
and continuous opinion ($-1 \le O_{i}(t)\le 1$).
The opinions of the two agents  
 are modified simultaneously. 
After $N$ such interactions, one  time step  is said to be completed. 
The  interacting agents are  chosen randomly from the $N$ agents and 
 thus one may consider the topology of the system to be like a fully connected graph.
%to be like a dynamic random graph where the connections are redefined at every step. 

 The behavior of the model was shown to be independent of the distribution from which $\mu_{ij}$ are drawn;
in the present work, we consider $\mu_{ij}=\pm 1$. 
If the opinion of an agent becomes 
higher (lower) than $+1(-1)$ following an interaction, then it is made equal to $+1(-1)$
 for both continuous and discrete opinions.
If the discrete opinions are taken as  $0$ and $\pm 1$ initially, 
$\mu_{ij} =\pm 1$ ensures that at subsequent times the opinion values will take up one of these values only.

In  models which are defined through dynamical rules, the question of equilibration may not be relevant and hence 
one introduces a damage in the beginning only  
in the traditional method (TM). 
We simulate two systems of $N$ individuals using the same initial
random discrete/continuous opinions except for one randomly chosen individual. Then the two systems are
allowed to evolve using the same random numbers. 
In the other (NVN) method, 
the initial systems are identical but different random numbers are used  
in the time evolution. This implies that the agents who interact in the two 
replicas are in general different in the NVN method.

The damage at time $t$ is defined as
\begin{equation}
\label{eq:cita}
\langle D(t) \rangle = \frac{1}{N}{\sum_i {|{O_{i}^{(1)}(t)}-O_{i}^{(2)}}(t)|},
\end{equation}
where $O_i^{(1)}(i)$ and  $O_i^{(2)}(i)$ are the $i$th agent's 
opinion in the two replicas.
We also calculate $P_{D}(t)$, the fraction of agents for which $O_{i}^{(1)} \neq O_{i}^{(2)}$.

In this model a parameter $q$ is used which denotes the fraction of negative interaction  
($\mu_{ij} = -1$). It was found in \cite{biswas} that below a particular value $q=q_c = 0.25$, the
system becomes ordered, while a disordered phase exists
for higher values of $q$. For $q \geq q_c$ the distributions of opinions are symmetric in both discrete and continuous cases. 
For discrete opinions, above $q_c$,   all three fractions of opinions ($0,\pm1$) are equally probable with probability $1/3$. 
For the continuous case also, all opinions $-1 \le O_{i}(t)\le 1$ are equally probable in the disordered phase, barring $O_i=0,\pm 1$. 
$O(i) = \pm1 $  has a higher 
probability compared to other values due to the imposed boundary condition.
When the interaction $\mu_{ij}$ between two agents particular agents $i$ and $j$ is kept unchanged throughout the time evolution, we call it quenched randomness.
On the other hand,  when a new value of $\mu_{ij}$ is chosen  every time the two agents interact, it is a case of annealed randomness.
%$\mu$ has been used  both as a $annealed$ variable which changes with time and $quenched$ variable in which case it is independent of time. 
It was shown in \cite{biswas} that the order-disorder phase transition which 
occurs with mean field critical behavior is independent of this choice.

In the simulation, we have taken systems with size $2^{8} \le N \le 2^{12}$. 
 For the same system, 400 different choices of initial random opinion of agents have been taken and
quantities are averaged over all the configurations. 

\section{RESULTS AND ANALYSIS}
\subsection{Results for the Traditional Method}
\subsubsection{Discrete opinion}

%111111111111111111111111111111111111111111111111111111111111
\begin{figure}[!h]
%\begin{center}
%\resizebox{70mm}{!}{\includegraphics[scale=1.2]{ep_col_del_0.7.eps}}
\resizebox{90mm}{!}{\includegraphics {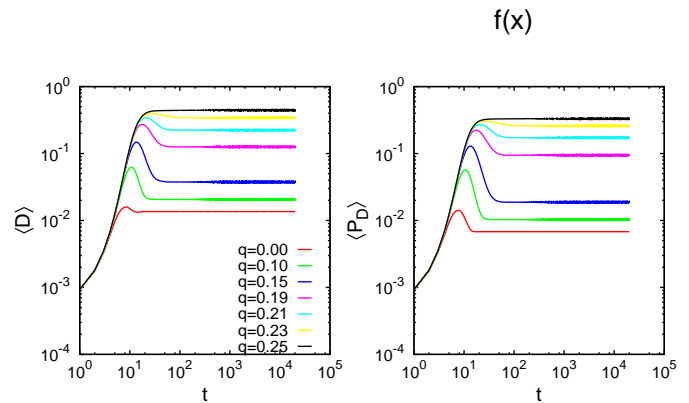}}
%\resizebox{120mm}{!}{\includegraphics{fig2.eps}}
%\end{center}
\caption{(Color online) (Traditional method (TM); quenched randomness) Plot of the average damage (left panel) and 
fraction of damaged agents (right panel) as a function of time for different values of $q$ for discrete opinion. All
data are for $N = 2048$.}
\label{Dyn_adis}
\end{figure}
%111111111111111111111111111111111111111111111111111111111111
%22222222222222222222222222222222222222222222222222222
\begin{figure}[!h]
%\begin{center}
%\resizebox{70mm}{!}{\includegraphics[scale=1.2]{ep_col_del_0.7.eps}}
\resizebox{90mm}{!}{\includegraphics {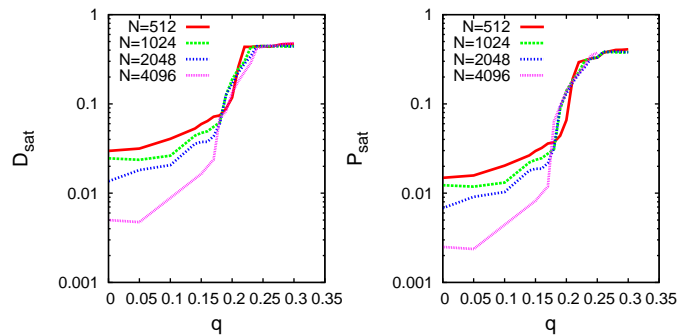}}
%\resizebox{120mm}{!}{\includegraphics{fig2.eps}}
%\end{center}
\caption{(Color online) (TM; quenched randomness) Variation of saturation value of damage (left panel) and fraction of damaged agents (right panel) with $q$ for discrete opinion.}
\label{sat_adis}
\end{figure}
%22222222222222222222222222222222222222222222222222222

%Average number of matched sites:
We first discuss the case when $\mu$ is chosen randomly in a quenched  manner.
 For the discrete opinion case the initial damage is introduced in one randomly chosen agent say, X, in
the following way: if the opinion $O_X$ of this  agent  is $1$ in replica $A$, then  the opinion in replica $B$ becomes $0$, otherwise it is ($O_X+1$).                                                                                                                                                                              
We calculate average damage $D(t)$ and fraction of average damaged agents $P_D(t)$ as  functions of time (Fig.~{\ref{Dyn_adis}).
 For the first few time steps both $D(t)$ and $P_{D}(t)$ 
increase sharply and then decrease for small values of $q$.
Finally $D(t)$ ($P_{D}(t)$) reaches a saturation value $D_{sat}$ ($P_{sat}$) which depends on $q$ up to $q \approx 0.25$.
The saturation value $D_{sat}$ is nonzero for all $q$. We plot the saturation values for different   system size $N$ as  functions of 
$q$ (Fig.~{\ref{sat_adis}).
Up to $q_d \approx 0.18$, $D_{sat}$ and $P_{Sat}$ decrease with $N$, above $q_d$ they show system size independent behavior. 
For $q \ge q_c = 0.25$, $D_{sat}$ and  $P_{sat}$ are independent of both $q$ and $N$.  

The results for the case when $\mu$ is considered as a annealed  random variable are similar (Figs~{\ref{Dyn_qdis} and {\ref{sat_qdis}). 
However, the saturation values below $q_d$ appear to be almost independent of $q$ which is in contrast to the quenched case.
Both $D$ and $P_d$ also show more pronounced non-monotonic behavior against time  close to $q_d$.

It is interesting to note that even for $q=0$, there is a small non-zero value of damage in finite systems. We will discuss this issue later in this section.

%333333333333333333333333333333333333333333333333333333333333
\begin{figure}[!h]
%\begin{center}
%\resizebox{70mm}{!}{\includegraphics[scale=1.2]{ep_col_del_0.7.eps}}
\resizebox{90mm}{!}{\includegraphics {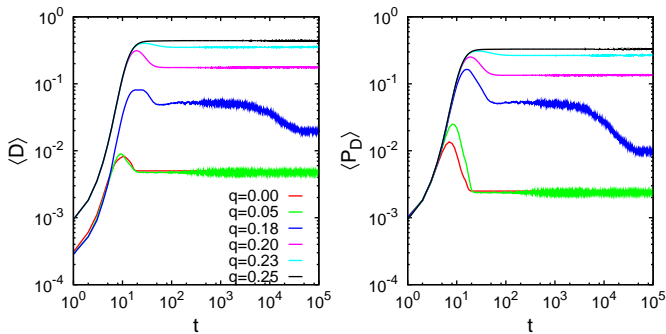}}
%\resizebox{120mm}{!}{\includegraphics{fig2.eps}}
%\end{center}
\caption{(Color online)(TM; annealed randomness) Plot of the average damage (left panel) and 
fraction of damaged agents (right panel) as a function of time for different values of $q$ for discrete opinion. All
data are for $N = 2048$.}
\label{Dyn_qdis}
\end{figure}

%444444444444444444444444444444444444444444444444444

\begin{figure}[!h]
%\begin{center}
%\resizebox{70mm}{!}{\includegraphics[scale=1.2]{ep_col_del_0.7.eps}}
\resizebox{90mm}{!}{\includegraphics {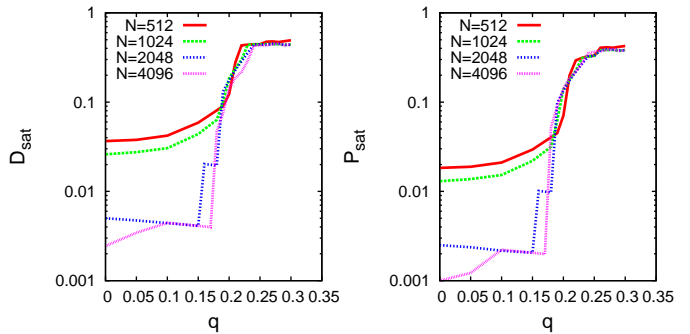}}
%\resizebox{120mm}{!}{\includegraphics{fig2.eps}}
%\end{center}
\caption{(Color online) (TM; annealed randomness) Variation of saturation value of damage (left panel) and fraction of damaged agents (right panel) 
with $q$ for discrete opinion.}
\label{sat_qdis}
\end{figure}

%saturation saturation saturation

\subsubsection{Continuous opinion}
In the continuous opinion case the results do not depend qualitatively and quantitatively on the manner in which $\mu$ are chosen (annealed or quenched).
Here the initial difference between two opinions of a single agent in the two replicas is taken as $0.01$. 
In this case, we take two opinions to be equal if their difference is less than $10^{-6}$. Also, the results do not depend qualitatively on the initial damage.
 It is observed that  below a certain value of $q$,
the value of $D(t)$ and $P_D(t)$ go sharply to zero after an initial increase and beyond this value of $q$, both reach a finite saturation value (Fig.~{\ref{Dyn_qcon}). 
We conclude that this  value of $q$ corresponds to 
 the  damage spreading transition point but it has a finite size dependence and therefore is denoted as $q_d (N)$.
%An estimate of $q_d$ is made by extrapolating the values obtained for different system sizes $N$.
 Plotting $q_d(N)$ versus $1/N$ (Fig.~{\ref{nonzero}}),  $q_d(N \to \infty$) is obtained as $\sim 0.17$ in the thermodynamic limit.
For $q>q_d$, $D_{sat}$ and $P_{sat}$ show nominal system size dependence (which is not systematic) and for $q\ge q_c$ these are also independent of $q$ (Fig.~{\ref{sat_qcon}}).

%555555555555555555555555555555555555

\begin{figure}[!h]
%\begin{center}
%\resizebox{70mm}{!}{\includegraphics[scale=1.2]{ep_col_del_0.7.eps}}
\resizebox{90mm}{!}{\includegraphics {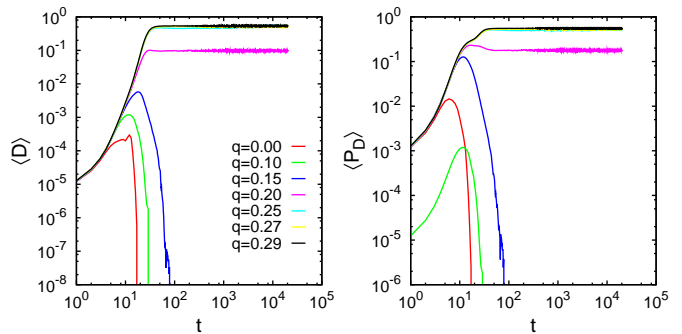}}
%\resizebox{120mm}{!}{\includegraphics{fig2.eps}}
%\end{center}
\caption{(Color online) (TM; quenched randomness) Plot of the average damage (left panel) and 
fraction of damaged agents (right panel) as a function of time for different values of $q$ for continuous opinion. All
data are for $N = 2048$.}
\label{Dyn_qcon}
\end{figure}
%666666666666666666666666666666
\begin{figure}[!h]
%\begin{center}
%\resizebox{70mm}{!}{\includegraphics[scale=1.2]{ep_col_del_0.7.eps}}
\resizebox{90mm}{!}{\includegraphics {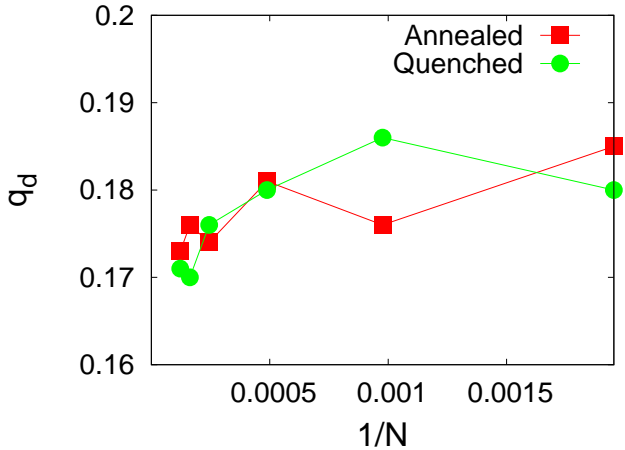}}
%\resizebox{120mm}{!}{\includegraphics{fig2.eps}}
%\end{center}
\caption{(Color online) Variation of $q_d$ with $1/N$ for continuous opinion using TM along with the fitted lines.}
\label{nonzero}
\end{figure}

%777777777777777777777777777777777777
\begin{figure}[!h]
%\begin{center}
%\resizebox{70mm}{!}{\includegraphics[scale=1.2]{ep_col_del_0.7.eps}}
\resizebox{90mm}{!}{\includegraphics {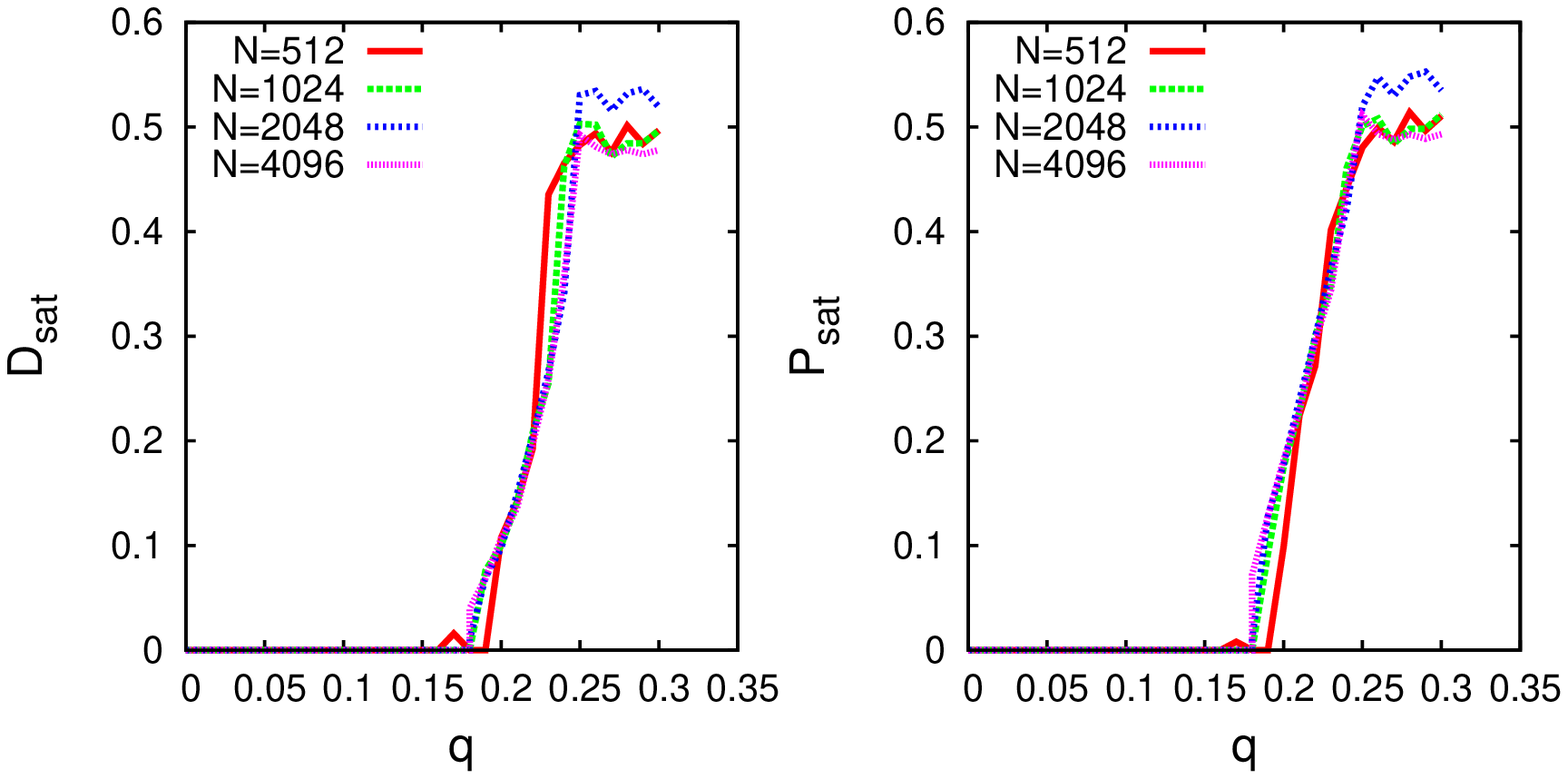}}
%\resizebox{120mm}{!}{\includegraphics{fig2.eps}}
%\end{center}
\caption{(Color online) (TM; quenched randomness) Variation of saturation value of damage (left panel) and fraction of damaged agents (right panel)  with $q$ for continuous opinion.}
\label{sat_qcon}
\end{figure}
%%%%%%%%%%%%%%%%%%%%%%%%%%%%%%%%%%%%%%%%%%%%%%%%%%%%%%%%%%%%%%%%%%%%%%%%%%%%%%%%%%%%%%%%%

\subsection{Nature versus nurture method} 
\begin{figure}[!h]
\resizebox{90mm}{!}{\includegraphics {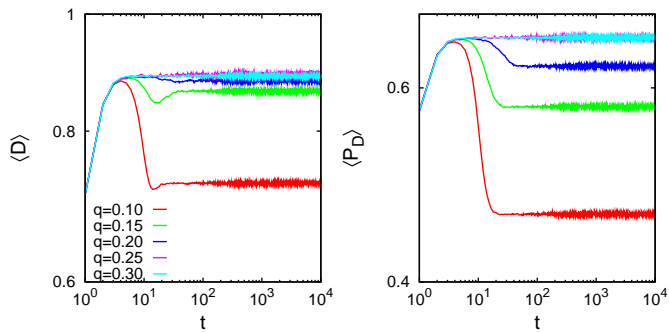}}
\caption{(Color online) (Nature versus nurture (NVN) method;  quenched randomness) Plot of the average damage (left panel) and 
fraction of damaged agents (right panel) as a function of time for different values of $q$ for discrete opinion. All
data are for $N = 2048$.}
\label{ddnvn}
\end{figure}
\begin{figure}[!h]
\resizebox{90mm}{!}{\includegraphics {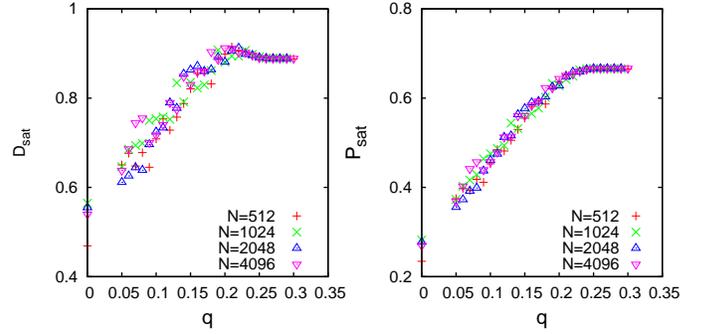}}
\caption{(Color online) (NVN; quenched randomness) Variation of saturation value of damage (left panel) and fraction of damaged agents (right panel) with $q$ for discrete opinion.}
\label{dsnvn}
\end{figure}
%Having observed that in the traditional method, the results are independent 
%of the nature of randomness and whether the opinions are discrete or continuous,for the NVN method, we consider only the quenched randomness case with discrete
%values of the opinion.

%We consider the variation of of .....
Having observed in the  TM that the nature of randomness does not affect the results significantly, we have
considered only  quenched randomness in $\mu_{ij}$ when using the NVN method (this involves less time to get the results computationally).

\subsubsection{Discrete opinions}

 Initially a random configuration with opinion  $\pm 1,0$ is chosen.
Starting with two  identical configurations and allowing them to evolve independently, we observe that the time dependence
of $D(t)$ and $P_D(t)$ is qualitatively similar to that in the TM (Fig.~{\ref{ddnvn}}). However, when saturation
 values are considered, we find that both $D_{sat}$ and $P_{Sat}$ show no appreciable systematic dependence 
on $N$; even for $q \to  0$, both remain finite  (Fig.~{\ref{dsnvn}}). 
Thus we conclude that the damage spreading transition occurs at $q_d = 0$ 
in this case. There is an increase with $q$ before  $D_{sat}$ or $P_{Sat}$ reaches 
a constant value close to $q\approx 0.25$.

\subsubsection{Continuous  opinions}

For the continuous opinion, a similar procedure is followed. 
In this case, like the TM,  we take two opinions to be equal if their difference is less than $10^{-6}$. Time evolution of $D(t)$ and $P_D(t)$ 
is shown in Fig.~{\ref{cdnvn}}. Both saturation values  are finite for all $q>0$  and show system size independent behavior.
For  $q \gtrsim 0.25$ saturation value of $D(t)$ and $P_D(t)$ show 
a $q$ independent behavior  (Fig.~{\ref{csnvn}}).

For small $q$,   both $D_{sat}$ and $P_{sat}$ are  order of magnitude smaller than that obtained in
 the discrete case, being $  {\mathcal{O}}(10^{-2})$. This is similar to the TM result.
However, in TM, the saturation values are zero up to a finite value of $q$. 
Here in contrast, $D_{sat}$ and $P_{sat}$ grow from  zero at 
$q=0$ itself with no system size dependence. So here too, the damage spreading
transition occurs at $q_d=0$ as in the discrete case.

\begin{figure}[!h]
\resizebox{90mm}{!}{\includegraphics {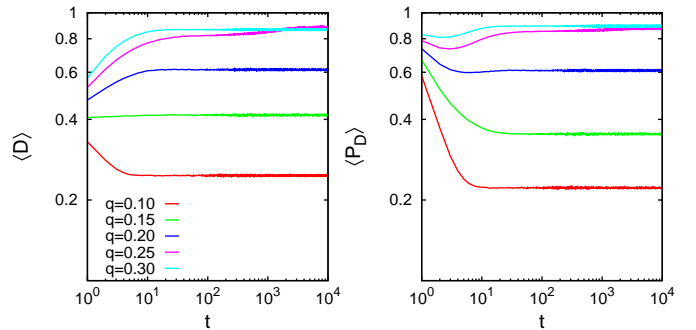}}
\caption{(Color online) (NVN; quenched randomness) Plot of the average damage (left panel) and 
fraction of damaged agents (right panel) as a function of time for different values of $q$ for continuous opinion. All
data are for $N = 2048$.}
\label{cdnvn}
\end{figure}
\begin{figure}[!h]
\resizebox{90mm}{!}{\includegraphics {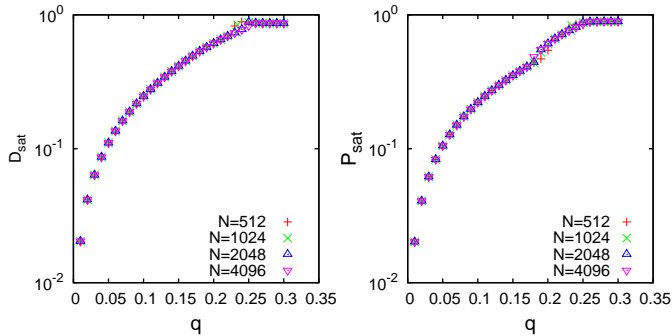}}
\caption{(Color online) (NVN; quenched randomness) Variation of saturation value of damage (left panel) and fraction of damaged agents (right panel) with $q$ for continuous opinion.}
\label{csnvn}
\end{figure}

\subsection{Theoretical estimates and comparison}
%Average number of damaged agents:
One can easily estimate $D_{sat}$ and $P_{sat}$ for $q>0.25$ theoretically when the system becomes completely disordered.
We denote the average damage and the fraction of 
disagreeing agents obtained   theoretically  by  $D_{est}$ and $P_{est}$ respectively.

For discrete opinion,  above $q=0.25$, all three types of opinion $(0,\pm 1$) have equal probability ($=1/3$) of occurrence~\cite{biswas}.
So one can estimate $P_{sat}$ and $D_{sat}$ for two configurations which are completely uncorrelated (which will happen ideally if the damage spreads through the entire system). 
%Assuming complete randomness the opinions of two agents can be distributed in nine ways and the ways are 
%$(-1,-1)$, $(-1,0)$, $(-1.1)$, $(0,-1)$, $(0,0)$, $(0,1)$, $(1,-1)$, $(1,0)$ and $(1,1)$. 
%In which the number of matched  cases are $(-1,-1)$, $(0,0)$ and $(1,1)$. 
%The average number of matched agent is $1/3$. 
%{Average damage:}
%Absolute value of difference of opinions (damage) between two agents in six unmatched cases 
%are $1$ ($(-1,0)$, $(0,-1)$), $1$ ($(1,0)$, $(0,1)$) and $2$ ($(1,-1)$, $(-1,1)$).
%Average damage assuming complete randomness is therefore $8/9$.
Assuming there is no correlation, the opinion of an agent in replica A and replica B can
have equal probability of having the following nine possible set of values:
 $(-1,-1), (-1,0),(-1,+1), (0,-1), (0,0), (0,+1), (+1,-1),\\ 
 (+1,0)$ and $(+1,+1)$.
Thus the average damage
\begin{eqnarray*}
 D_{est} = (2\times2/9+4\times1/9) = 8/9.
\end{eqnarray*}
Also the probability $P$ that the opinions are different is equal to $P_{est}=2/3$. 

For continuous opinions, the probability that
the opinions are exactly equal in the two replicas would be zero for completely uncorrelated 
replicas and $P_{est}=1$. Denoting the opinion distribution by ${\mathcal P(x)}$, the square of average damage can be calculated as
\begin{eqnarray*}
D_{est}^2 &=&{\frac { \int_{-1}^1 \! \int_{-1}^1 (x_{1}-x_{2})^2 {\mathcal P(x_1)} {\mathcal P(x_2)}\,dx_{1}\,dx_{2}\ }{ \int_{-1}^1 \! \int_{-1}^1 {\mathcal P(x_1)} {\mathcal P(x_2)}\,dx_{1}\,dx_{2}\ }},\\
%&= &{\frac { \int_{-1}^1 \! \int_{-1}^1 (x_{1}^2+x_{2}^2) P(x_1) P(x_2)\,dx_{1}\,dx_{2}\ }{ \int_{-1}^1 P(x_1) dx_{1}\ \! \int_{-1}^1  P(x_2)\,dx_{2}\ }}\\
%&=&{\frac {2.\frac{2}{3}}{2.2}}\\
&= &{\frac {2}{3}},
\end{eqnarray*}
assuming ${\mathcal P(x)}$  to be uniform above $q_c$.
Therefore the expected value of average $D_{est}=\sqrt{2/3} \approx 0.82$
%\subsubsection{Continuous opinion}

%Average number of matched sites:

%\begin{eqnarray}
%N(x)&=&\frac { \int_{-1}^1 \! \int_{-1}^1 P(x_1) P(x_2)\,dx_{1}\,dx_{2}\ \delta(x_{1}-x_{2})}{ \int_{-1}^1 \! \int_{-1}^1 P(x_1) P(x_2)\,dx_{1}\,dx_{2}\ }\\
%&= &\frac { \int_{-1}^1 \ [(P(x_1)]^2 dx_{1}}{ \int_{-1}^1 P(x_1) dx_{1}\ \! \int_{-1}^1  P(x_2)\,dx_{2}\ }\\
%&= &\frac {2}{2.2}\\
%&= &\frac {1}{2}
%\end{eqnarray}
%For quasi-continuous case, $x$ takes values $1, 2, 3, ........N$. For two completely random configurations, probability that each equal to each is $(1/N)(1/N)N= 1/N$.
%For $N \to \infty$, this goes to zero. Therefore the fraction of sites for which opinions disagree should be $1$.

%\vspace{3cm}

%[Average Damage]$^2$:
%\begin{eqnarray}
%D^2 &=&{\frac { \int_{-1}^1 \! \int_{-1}^1 (x_{1}-x_{2})^2 P(x_1) P(x_2)\,dx_{1}\,dx_{2}\ }{ \int_{-1}^1 \! \int_{-1}^1 P(x_1) P(x_2)\,dx_{1}\,dx_{2}\ }}\\
%&= &{\frac { \int_{-1}^1 \! \int_{-1}^1 (x_{1}^2+x_{2}^2) P(x_1) P(x_2)\,dx_{1}\,dx_{2}\ }{ \int_{-1}^1 P(x_1) dx_{1}\ \! \int_{-1}^1  P(x_2)\,dx_{2}\ }}\\
%&=&{\frac {2.\frac{2}{3}}{2.2}}\\
%&= &{\frac {2}{3}}
%\end{eqnarray}
%Therefore $D=0.8164$ and
%expected value of average damage is $\approx 0.8164$.

\subsubsection{Comparison with TM} 

The observed values of  $P_{sat}$ and $D_{sat}$ for large $q$ ($D_{sat} \sim 0.47,  
P_{sat} \sim 0.39 $ and $D_{sat} \sim 0.50,  
P_{sat}\sim 0.50$ for the discrete and continuous opinions respectively) for TM 
however do not match with the theoretical estimates.
For both the discrete and continuous cases, the values of $P_{sat}$ and $D_{sat}$ are far less than $P_{est}$ and $D_{est}$.
This shows that the damage must have spread only partially (or not at all) in a 
finite fraction of cases. We study this in detail in the next subsection by calculating the distribution of $P_{sat}$.
In case $P_{sat}<P_{est}$, one can say there is positive correlation between the two replicas, otherwise there is negative correlation.
The latter is possible only for the discrete opinion case as for the continuous opinions, $P_{est}=1$.

\subsubsection{Comparison with NVN method}

In comparison, we note that in the NVN method, the saturation values  of $P_D$ and $D$
for $q>0.25$ coincide with the estimated values quite well. 
Thus the disordered states are simply uncorrelated here, all initial
correlation is completely destroyed. 
This is in stark contrast to the result of the traditional method.
%which shows that even at large values of $q$, there is a finite probability 
%that the initially damaged configurations reach highly correlated states. 

%88888888888888888888888888888888888
\subsection{Distribution of fraction of damaged site in TM}

In the TM, we noted that the estimated saturation values of  $D$ and $P$  are clearly different from 
the observed ones for $q > q_c$. 
This may happen due to two reasons. Either the systems actually evolve to correlated configurations (conjecture 1), 
or there may be some configurations for which they reach identical or nearly identical states and
some for which they evolve to uncorrelated states (conjecture 2). 
In order to check this, we conduct a systematic study of the probability distribution $R(P_{sat})$ for all $q$ values.  

For the discrete case, $R(P_{sat})$ shows an interesting behavior even when $q$ is quite small.
 It has non-zero values at $P_{sat} = 0$ and at a fairly large value of $P_{sat}$ close to unity. This signifies that at small values of $q$,
when the system is almost fully ordered (almost all opinions equal to 
+1 or -1 \cite{biswas}), the initial damage leads in a few cases to configurations 
with strong negative correlations. For the extreme case $q=0$, these 
two configurations correspond to the   all +1  all -1  states.
However, as previously noted, this effect vanishes as the system size becomes larger. 
In continuous opinions, no such phenomenon is observed. 

In general, for $q >> q_d$, $R(P_{sat})$  has a bimodal nature having non-zero values for $P_{sat} =0$ and a few larger values of $P_{sat}$ close to $P_{est}$.  
We indeed find that even for large values of $q$, $R(P_{sat}=0)$ is non zero which explains the discrepancy between the observed 
and theoretical results 
and indicates that the second conjecture is true. 
% $R(P_{sat})\ne 0$ even for $P_{sat}>P_{est}$ in the discrete case which shows for very large $q$ values negative correlation
%have developed here. 
For the continuous case, one gets $R(P_{sat}=1)=0$ even when $q$ is very large as the high density of 
agents with opinions equal to $\pm 1$ induce some positive correlation (not considered in the theoretical estimates).
% We find that in general the distribution is bimodal, there being a peak at $N_D=0$ and another at a
%nonzero value of $N_D$ which depends on $q$. The second peak occurs close to the theoretically estimated values for large
%values of $q$ as expected.
The results are plotted in Figs~{\ref{Dist} and {\ref{Dist_zero}}.
$R(P_{sat}=0)$ has an interesting behavior; it is almost equal to $1$ 
upto $q_d$ and shows a sharp fall to another constant non-zero value beyond
$q_d$ in both discrete and continuous cases.  
The constant value of $R(P_{sat} = 0)$ for $q > q_d$ is about 0.5 and 0.4 for the discrete 
and continuous 
opinions respectively.

%saturation saturation saturation
\begin{figure}[!h]
%\begin{center}
\resizebox{90mm}{!}{\includegraphics {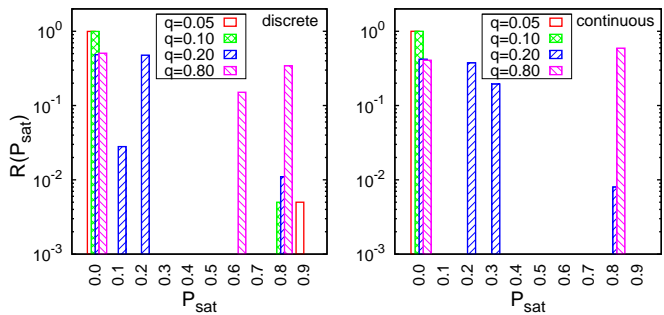}}
%\resizebox{120mm}{!}{\includegraphics{fig2.eps}}
%\end{center}
\caption{(Color online)  Histogram showing   $R(P_{sat})$, 
the probability of  fraction of damaged agents for  different values of $q$
using TM. 
 Data are for $N = 2048$.}
\label{Dist}
\end{figure}

%%%999999999999999999999999999999999999999999999

\begin{figure}[!h]
\begin{center}
\resizebox{70mm}{!}{\includegraphics {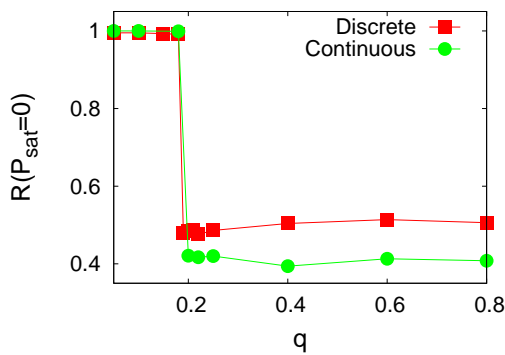}}
\end{center}
\caption{(Color online)  Variation of probability of zero damage with $q$ in TM. Data are for $N = 2048$.}
\label{Dist_zero}
\end{figure}

%
%
%
%%
%%%
%%%
%
%%
%%
%%
%%%
\section{Summary and Discussions}
We have studied the damage spreading phenomena in an opinion dynamics model which has a known order disorder transition
and where one can use both discrete and continuous opinions. 
We employ two different methods, the traditional method and 
the nature versus nurture method \cite{machta} to study the phenomena. In both cases, the dynamics of the damage $D$ shows a non-monotonicity 
which makes it difficult to comment on the exact nature of $D(t)$  as a function of time. Thus unlike in many other studies,
 no estimate of exponents associated with $D(t)$ can be made. 
This is perhaps a feature which makes this model different from 
conventional dynamical models. At large times, $D(t)$ reaches
 a saturation value $D_{sat}$. The behavior of $D_{sat}$ as a function of the model parameter $q$ reveals the existence of a damage
 spreading transition at $q=q_d$. We find that $q_d$ is less than $q_c$, the order disorder phase transition point.  
This is true for both the methods, in fact for NVN, $q_d =0$. 
%$D_{sat}$ attains finite values independent of system size above $q_d$. 
It is difficult to ensure the nature of transition, apparently $D_{sat}$
varies continuously with $q$ above $q_d$ before attaining a $q$ independent value for $q > q_c$ where the system enters the disordered phase.
However, there is no system size dependence of $D(t)$ and $P(t)$ above $q_d$ so that conventional finite size 
scaling analysis for continuous phase transition is also not possible to do. 

As mentioned before, in many physical systems, the question whether the damage spreading transition coincides with the order-disorder 
transition  point has been investigated.
Here we clearly get $q_d<q_c$.
The significance of this result in TM is that for $q_d < q < q_c$, even though 
consensus is reached, very small  changes in even a single agent may lead
to a different consensus  state with a finite probability.  
In NVN,  $q_d = 0$  implies that  if the same agent goes through a different sequence of 
interactions, the result will be  different for any $q$ 
with finite probability.  
The NVN result is consistent with the finding of \cite{machta} in the 
Ising model, where even at zero temperature, the overlap between 
identical states vanishes, albeit in a much slower manner. 
Above $q_c$,  
i.e.,  in the disordered state, even  when systems are slightly different initially, 
if they follow the same environment, they have a finite probability 
of  ending  up in  
configurations with large overlap while that is never true if the environment is completely 
different.  This shows that if the initial states are damaged and in addition
they are evolved independently, they will always lead to uncorrelated states. 
Although the damage spreading transition occurs before the order-disorder transition ($q_d<q_c$), 
we find that the damage is sensitive to the static critical point as it attains a 
saturation value for $q\geq q_c$. This is true in both methods.

%Damage spreading has been shown to be dependent on the precise dynamical rules in Ising model where one can use several dynamical schemes like 
%Glauber, Metropolis, Heat-bath, Kawasaki algorithm etc. 
We have considered different nature of randomness and 
opinions.  The nature of randomness apparently does not play any role. 
%The dynamical rule given by eq. ($1$) is uniquely defined but the randomness can be introduced in
%different ways. Choosing $\mu=1$ and $-1$ with probability $(1-q)$ and $q$ respectively, we use annealed and quenched values of $\mu$ while considering the interaction between two agents.
%This however does not affect the results. 
Rather, the results depend on the nature of the opinions. Not only is   the dynamical evolution 
  quite different for the two cases below $q_d$ in finite systems, 
the estimated damage spreading transition point also appears to be slightly 
different as noted in TM. For NVN too, we find the fluctuations to be larger in case of 
discrete opinions which, however, is not surprising. However, $q_d=0$ for both discrete and continuous in NVN.
The distribution of $P_{sat}$ is also different 
for the two types of opinions as checked in TM, in particular, for, $P_{sat} =0$, the values differ markedly for $q > q_d$. 
% Such differences did not appear in the  
%static critical 
%phenomena.  

One of the main issues in the present work is the comparison of the two cases of  damage spreading. Both qualitative and quantitative differences are noted
in the results. The transition occurs at $q_d = 0$ for NVN in contrast to a finite value of  $q_d$ obtained in TM. 
The overlap of the evolved states  above $q_c$ also shows 
an interesting difference. 
In the disordered phase ($q > q_c$), the expected value of $D_{sat}$ and $P_{sat}$ considering totally
uncorrelated opinions in the two time evolved  configurations 
are much larger compared to the  values obtained in TM but 
for NVN these are very close to the observed values. 
Further study of the distribution shows that in TM, there are 
configurations where damage does not spread at all leading to 
 values less than that expected for completely 
uncorrelated cases.

%The present study of damage spreading has led to a number of interesting results:  (a) the dependence of the dynamics on opinion type  (b) the existence of a phase transition
%(c) the non-percolation of damage in finite number of cases and  also that 
%the probability that the opinion of all agents match having a step function like behavior close to $q_d$.
%

%Time dependent properties

%Phase transition 

%Discrete/Continuous opinions 

%Damage spreading distribution 

%significance of results

\medskip

Acknowledgements: 
We thank Soumyajyoti  Biswas and Arnab
 Chatterjee for a critical reading of an earlier version of the manuscript and  useful comments. 
AK acknowledges financial support from UGC sanction no. UGC/534/JRF(RFSMS). 
PS is grateful to UPE (UGC) project. 
%AK acknowledges financial support from UGC sanction no. UGC/534/JRF(RFSMS).

\end{document}